\documentclass[aps,prl,showpacs,twocolumn,superscriptaddress,longbibliography,amsmath,amssymb]{revtex4-1}
\usepackage[pdftex]{graphicx}
\usepackage{color}
\usepackage[pdftex,bookmarks,colorlinks,breaklinks]{hyperref}
\hypersetup{linkcolor=red,citecolor=blue,filecolor=dullmagenta,urlcolor=blue}
\hypersetup{pdftoolbar=true,pdfpagemode=UseNone,pdfstartview=FitH, colorlinks=true,extension=}

\usepackage{braket}

\usepackage{soul}

\begin{document}

\title{{\it In-situ} Physical Adjoint Computing in multiple-scattering electromagnetic environments for wave control}  

\author{John Guillamon}
\altaffiliation{These authors contributed equally to this work}
\affiliation{Wave Transport in Complex Systems Lab, Department of Physics, Wesleyan University, Middletown, CT-06459, USA}
\author{Cheng-Zhen Wang}
\altaffiliation{These authors contributed equally to this work}
\affiliation{Wave Transport in Complex Systems Lab, Department of Physics, Wesleyan University, Middletown, CT-06459, USA}
\author{Zin Lin}
\affiliation{Bradley Department of Electrical and Computer Engineering, Virginia Tech, Blacksburg, VA 24060, USA}
\author{Tsampikos Kottos}
\email[]{tkottos@wesleyan.edu}
\affiliation{Wave Transport in Complex Systems Lab, Department of Physics, Wesleyan University, Middletown, CT-06459, USA}

\begin{abstract}
Controlling electromagnetic wave propagation in multiple scattering systems is a challenging endeavor due to the 
extraordinary sensitivity generated by strong multi-path contributions at any given location. Overcoming such 
complexity has emerged as a central research theme in recent years, motivated both by a wide range of applications 
-- from wireless communications and imaging to optical micromanipulations -- and by the fundamental principles 
underlying these efforts. Here, we show that an {\it in-situ} manipulation of the myriad scattering events, achieved 
through time- and energy-efficient adjoint optimization (AO) methodologies, enables {\it real time} wave-driven 
functionalities such as targeted channel emission, coherent perfect absorption, and camouflage. Our paradigm shift 
exploits the highly multi-path nature of these complex environments, where repeated wave-scattering dramatically 
amplifies small local AO-informed system variations. Our approach can be immediately applied to in-door wireless 
technologies and incorporated into diverse wave-based frameworks including imaging, power electronic and optical 
neural networks.
\end{abstract}

\maketitle


Controlling electromagnetic wave propagation in naturally occurring or engineered multi-mode complex media is a 
core challenge for RF/microwave, modern optical, and photonic systems \cite{cao2022shaping,gigan2022roadmap,Bliokh_2023,
carneiro2022study,de2021electromagnetic,anlage20,anlage22,jiang2024coherent,wang2024nonlinearity,goicoechea2024detecting,
sol2023reflectionless,popoff2014coherent,hsu2017correlation,chen2020perfect,bender22,mcintosh2024delivering,cizmar2015}. 
The origin of this difficulty lies in multiple scattering and the consequent interference of many photon paths, leading 
to extraordinary complexity and sensitivity in these media. Yet, controlling these wave-scattering events and their 
associated interference phenomena is essential for a wide range of applications, including satellite and in-door 
wireless communications, fiber-based communications and endoscopy, deep-tissue imaging, and optogenetic control of 
neurons. At first glance, the complete scrambling of a wavefront as it propagates through a complex medium 
appears to conflict with the objective of precision wave-control -- such as focusing electromagnetic radiation on 
a diffraction-limited spot inside or through a multi-scattering/opaque medium. Indeed, for many years, the presence 
of random secondary sources (scatterers or reflectors) was considered detrimental. However, novel techniques such as 
time-reversal \cite{Fink1,Fink2} and wavefront shaping (WS) \cite{Fink2,cao2022shaping,gigan2022roadmap} disrupted 
this paradigm by recognizing that these secondary sources offer additional degrees of freedom. Wavefront shaping 
protocols have relied on recent technological developments with spatial light/microwave modulators \cite{SLM1,Fink2,SMM1,SMM2}; 
these allow phase and/or amplitude modulation to each segment of an incident monochromatic wavefront in order to 
achieve desired functionalities after propagation through the complex medium. On the other hand, 
time-reversal provides a broadband approach that yields spatiotemporal focusing of waves. 

\begin{figure*}
\centering
\includegraphics[width=1.0\linewidth]{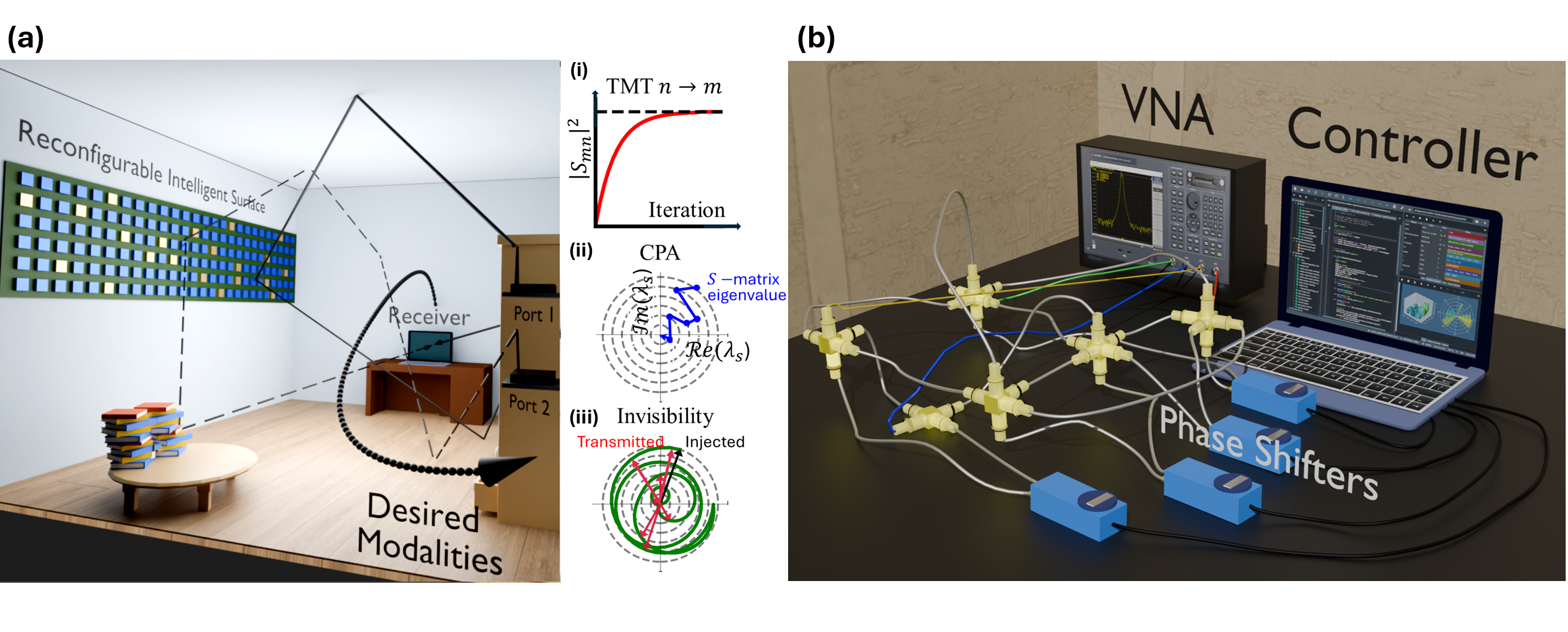}
\caption{ \label{Fig1}
{\bf In-door Wireless Communications and Experimental Platforms:} (a) A smart electromagnetic environment utilizing
RIS for a variety of modalities: (i) Targeted Mode Transfer that aims to receive the injected electromagnetic signal at 
a specific receiving channel; (ii) Coherent Perfect Absorption that aims to absorb the injected electromagnetic signal 
completely; (iii) Invisibility (cavity camouflaging) that results in an outgoing signal being the same (phase and amplitude) 
as the injected one. (b) A multi-resonant, multi-scattering complex network of coaxial cables has been used as a platform 
to demonstrate the viability of the iPAC protocol. The protocol used as control parameters, the relative amplitudes and 
phases of injected waves (which were controlled by the VNA), and a targeted set of coaxial cables of the network whose 
electrical lengths were digitally controlled using phase-shifters.
}
\end{figure*}

Although both of these methodologies guarantee optimal efficiency, they require a complete knowledge of the scattering 
domain, limiting their practicality for a variety of applications. A pivotal example is indoor wireless communications 
\cite{carneiro2022study,RIS3,RIS5}, where small temporal variations in the enclosure can drastically alter the scattering 
process. An entirely different approach relying on smart electromagnetic environments has emerged with the advent 
of reconfigurable intelligent metasurfaces (RISs) \cite{RIS1,RIS2,RIS3,RIS4,RIS5,RIS6,RIS7}. This approach foresees a 
fully programmable wave propagation to harness the wave-scattering complexity and achieve optimized transmission of both 
information and power. A bottleneck for the practical implementation of this proposal is the development of smart, 
low-cost/high-efficiency, optimization schemes that will be able to 
identify in real-time, with low latency, optimal RIS configurations for achieving specific modalities.

Meanwhile, there has been widespread attention towards physical (optical) analog computing for low-latency deep learning applications~ \cite{wetzstein2020inference,shastri2021photonics,shen2017deep,wright2022deep,roques2020heuristic,vadlamani2020physics,hughes2018training,
pai2023experimentally,xue2024fully}. Pioneering works such as~\cite{hughes2018training,pai2023experimentally} showed that it is 
possible to perform in-situ 
backpropagation through a photonic implementation of an artificial neural network. However, these platforms typically consist of 
\textit{feed-forward} waveguides, couplers and interferometers, in contrast to complex multi-scattering multi-resonant electromagnetic 
environments, in which \textit{real-time} optimizations of \textit{on-demand} wave-control functionalities are needed. Here, we 
experimentally demonstrate an {\it in-situ} Physical Adjoint Computing (iPAC) optimization protocol that leverages 
adjoint sensitivity analysis~\cite{cea1986conception,jameson1988aerodynamic,jensen2011topology,molesky2018inverse} to control 
and harness complex wave dynamics. The protocol is built around three components: in-situ measurements, targeted 
perturbations, and an external control mechanism, which collectively enables the real-time optimization of wave systems through 
two sequential field propagations -- forward and adjoint. First, local probes are employed to measure both the forward and adjoint 
wave fields at specific elements within the system. Forward propagation is utilized to compute a desired merit (or cost) function 
and to determine the excitation profile required for the subsequent adjoint propagation. The adjoint field, in turn, provides a 
\textit{comprehensive} and \textit{simultaneous} measurement of \textit{all} targeted sensitivities. An external control 
mechanism evaluates these sensitivities to identify the potential perturbations that could enhance (or diminish) the merit 
(cost) function. These adjustments are then delivered by local actuators to the targeted components (i.e., the tunable degrees). 
The cycle is repeated as many times as necessary to maximize (minimize) the merit (cost) function. To demonstrate 
the versatility of our protocol we showcase three different modalities, namely, targeted channel emission, coherent perfect 
absorption, and camouflaging, using a microwave experimental platform. The latter consists of a network of coupled coaxial 
microwave cables whose wave transport demonstrates features characterizing wave chaotic systems \cite{kottos2001,kottos2003c,
kottos2003b}. These networks are frequently used as models for mesoscopic quantum transport, sound propagation, and electromagnetic 
wave behavior in complex interconnected structures such as buildings, ships, and aircrafts \cite{hurt2000mathematical,kuchment2002graph,
kuchment2004quantum,berkolaiko2013introduction,tanner2022,wang2023} and therefore constitute a versatile platform for experimentally 
implementing our in-situ optimization protocol.

\section{Principles of In-Situ~Adjoint~Optimization}
Formally, the steady-state propagation of a time-harmonic wave field $\Phi$ is governed by a linear system: ${\cal M}({\bf p}) 
\Phi = {\bf b}({\bf p})$. Here, ${\cal M}$ is the system matrix, ${\bf b}$ is the driving source, and ${\bf p}$ is a vector of 
${\cal N}$-controllable optimization parameters. An optimization objective $g$ is typically expressed as an explicit function of $\Phi$ 
and ${\bf p}$, i.e., $g=g(\Phi,{\bf p})$. Using the chain rule, the gradient sensitivities of $g$ with respect to ${\bf p}$ are 
given by:
\begin{align}
    {d g \over d {\bf p}} = {\partial g \over \partial {\bf p}} + \left( {\partial g \over \partial \Phi}^T {\cal M}^{-1} 
    \left( {\partial {\bf b} \over \partial {\bf p}} - {\partial {\cal M} \over \partial {\bf p}} ~ \Phi  \right) \right)
\end{align}
Here, $ {\partial {\bf b} \over \partial {\bf p}} - {\partial {\cal M} \over \partial {\bf p}} ~ \Phi $ physically represents 
a collection of induced excitations resulting from perturbing the system via one parameter $p_i$ at a time for each $i=1,2,3,
\cdots,{\cal N}$. Consequently, $U = {\cal M}^{-1} \left( {\partial {\bf b} \over \partial {\bf p}} - {\partial {\cal M} \over 
\partial {\bf p}} \, \Phi  \right)$ denotes a collection of \textit{several} wave fields in response to each and every one of 
these perturbations (the $i$th column, $U_i$, corresponds to the wave field generated by perturbing the single $p_i$). However, 
finding the entire $U$ becomes excessive especially when the number of controllable parameters, $N$, is large.

The adjoint method addresses this challenge by reformulating the problem as:
\begin{align}
    \left( {\partial g \over \partial \Phi} \right)^T {\cal M}^{-1} = \Psi^T \Rightarrow 
    {\cal M}^T \Psi = {\partial g \over \partial \Phi}
\label{backward}
\end{align}
Here, $\Psi$ is the adjoint field generated in response to the source ${\partial g \over \partial \Phi}$. For {\it reciprocal} 
wave media, where ${\cal M}^T = {\cal M}$, the adjoint field can be found by propagating through the same system. The gradient 
sensitivities are now given by:
\begin{align}
    {d g \over d {\bf p}} = {\partial g \over \partial {\bf p}} +  \Psi^T \left( {\partial {\bf b} \over \partial {\bf p}} - 
    {\partial {\cal M} \over \partial {\bf p}} ~ \Phi  \right) \label{eq:adj}
\end{align}
This formulation significantly reduces computational demands, as all sensitivities can be obtained {\it through a single additional 
field propagation, instead of computing the entire collection $U$ of $N$ wave fields}. Moreover, ${\partial {\bf b} \over \partial 
{\bf p}}$ and ${\partial {\cal M} \over \partial {\bf p}}$ are typically very sparse tensors since the effect of each parameter 
$p_i$ on ${\cal M}$ and ${\bf b}$ is localized. Consequently, only the values of $\Phi$ and $\Psi$ corresponding to the non-zero 
entries of ${\partial {\bf b} \over \partial {\bf p}}$ and ${\partial {\cal M} \over \partial {\bf p}}$ are needed. 

To implement the adjoint method experimentally, we sequentially excite the wave system with the driving sources ${\bf b}$ and 
${\partial g \over \partial \Phi}$, measure $\Phi$ and $\Psi$ at the strategic positions designated by ${\partial {\bf b} \over 
\partial {\bf p}}$ and ${\partial {\cal M} \over \partial {\bf p}}$ and then compute ${d g \over d {\bf p}}$ digitally using 
Eq.~(\ref{eq:adj}). Importantly, our in-situ protocol bypasses the computationally intensive tasks of solving ${\cal M} \Phi 
= {\bf b}$ and ${\cal M} \Psi = {\partial g \over \partial \Phi}$. Instead, we directly measure $\Phi$ and $\Psi$, inherently 
accounting for all the complexities of the system, including hidden losses and detunings, thereby enabling self-calibration. 
Having found ${d g \over d {\bf p}}$, any gradient-guided optimization algorithm can be applied to advance $g$~\cite{NLopt}. 
We set up an external control enclosure to orchestrate the entire process, including the sequential (forward and adjoint) 
wave-field excitations, in-situ measurements, gradient computations, and optimization updates, ensuring seamless and efficient 
real-time optimization. 

While adjoint analysis shares a conceptual common ground with the celebrated backpropagation algorithm, our goal is not to develop 
a physical deep-learning platform~\cite{hughes2018training}. Instead, we aim to optimize a wave system in-situ to achieve specific physical functionalities in real time, 
such as perfect absorption, signal delivery to targeted channels, or camouflage. Unlike data-driven methods, our protocol does not 
train any neural network nor rely on extensive datasets. Crucially, our work should be distinguished from physical implementations 
of \textit{feed-forward} neural networks~\cite{hughes2018training,pai2023experimentally}, which often avoid back reflections. 
In contrast, our in-situ optimization holistically exploits the intricate physics of multiple scattering of waves within an arbitrarily 
complex network topology, where \textit{any} wave effect, including back reflections and even resonant phenomena, can be utilized 
as valuable physical degrees of freedom. Furthermore, our implementation at RF and microwave frequencies allows us to easily access 
both phase and amplitude information of the fields, which ensures that the intricate wave interactions within the system are accurately 
accounted for, enabling precise and reliable optimization.

\section{Physical Platform and Implementation of the Adjoint Protocol}

The complex microwave network \cite{Kottos1997,kottos2000,kottos2003c,sirko,wang2024bound,chen2020perfect} consists of $n=1,\cdots,V$ 
vertices, that are connected by one-dimensional coaxial wires (bonds) $B=(n,m)$ of length $L_B$, which are irrationally related to one 
another. The position $x_B=x$ on bond $B$ is $x=0 \,(l_B)$ on vertex $n (m)$. The connectivity of the network is encoded in the $V 
\times V$ symmetric adjacency matrix ${\cal A}$, with elements ${\cal A}_{nm} = 1$ if vertices $n\neq m$ are connected via a bond 
$l_B$, and ${\cal A}_{nm} = 0$ otherwise. The electric potential difference (voltage) between the inner and outer conductor surfaces 
of the coax cables at position $x$ along each bond satisfy the telegraph equation \cite{Kottos1997,kottos2000,kottos2003c,sirko,
wang2024bound} 
\begin{equation}
    \left(\frac{d^2 }{dx_{B}^2} + k^2\right)\psi_{B}\left(x_{B}\right) = 0;\quad k=\frac{\omega n_r}{c} 
    \label{telegraph}
\end{equation}
where $k$ is the wavenumber of the propagating wave with angular frequency $\omega$, $c$ is the speed of light in vacuum and $n_r$ is 
the complex-valued relative index of refraction of the coaxial cable whose imaginary part describes Ohmic losses in the cables. To 
emulate realistic conditions, we have considered that all cables suffer Ohmic losses which are modeled by a complex refractive index 
with imaginary part ${\cal I}m(n_r)\simeq 0.0085$. Furthermore, it is convenient to define the vertex field $\Phi = \left( \phi_1, 
\phi_2, \dots, \phi_N \right)^\top$ where $\psi_{B}\left(x_{B} = 0\right) = \phi_n$ is the voltage at vertex $n$. 

The scattering set-up is completed by connecting $\alpha=1,\cdots,N\leq V$ of the vertices to transmission lines (TL) that are used to 
inject and receive monochromatic waves of angular frequency $\omega=2\pi f$. The coupling to the TLs is described by the $N\times V$ 
matrix $W$ with elements $1 \, (0)$ when a vertex is connected (not connected) to a TL. At each vertex $n$, the continuity of the field 
and current conservation are satisfied. In the frequency domain, these conditions take the following compact form \cite{kottos2000,kottos2003c}
\begin{equation}
    \left( H(k) + i W^{\top} W \right) \Phi = {\bf b}; H_{nm} = \begin{cases}
        -{\displaystyle \sum_{l \neq n} }{\cal A}_{nl} \cot\left( k L_{nl} \right), & n = m, \\
        {\cal A}_{nm}\, \csc\left( k L_{nm} \right), & n \neq m.
    \end{cases}
    \label{grapheqmot}
\end{equation}
Above, ${\bf b} = 2 i W^{\top} {\bf {\it I}}$ is the $N$-dimensional vector that describes the driving source, and $I_\alpha= A_{\alpha}
e^{i\theta_{\alpha}}$ are the components of an $L$-dimensional vector that describes the amplitudes $A_{\alpha}$ and the phases 
$\theta_{\alpha}$ of the input fields $I_{\alpha}$ from the $\alpha$-lead. Consequently, the system matrix for the microwave network 
is ${\cal M}= \left( H(k) + i W^{\top} W \right)={\cal M}^\top$.

The gradient sensitivities are evaluated using Eq. (\ref{eq:adj}). The implementation of this equation requires the knowledge of the 
adjoint field $\Psi$ which is the solution of the adjoint Eq. (\ref{backward}). In our case, it takes the same form as the equations 
that dictate the forward field with the only difference being the driving source vector, i.e., ${\cal M} \Psi = 2 i W^{\top} \left( 
\frac{\partial g}{\partial \Phi} \right)^{\top}$. The latter is determined from the specific form of the optimization objective function 
$g(\Phi)$. 

The other elements required for the evaluation of Eq. (\ref{eq:adj}) are the gradients ${\partial {\cal M} \over \partial {\bf p}}$ 
and ${\partial {\bf b} \over \partial {\bf p}}$. The optimization parameter vector ${\bf p}$ is partitioned into two parts: the first 
one involves cavity-shaping optimization parameters (e.g. selected set of bond lengths $\{L_{nm}^{\rm opt}\}$ in the network), which are 
encoded in the system matrix ${\cal M}$. Its gradient $\frac{\partial {\cal M}}{\partial L_{nm}^{\rm opt}}$ is a sparse $V \times V$ 
operator with non-zero elements only at entries that incorporate the selected bonds $\{L_{nm}^{\rm opt}\}$. We also consider additional 
optimization parameters, i.e., the amplitudes $A_\alpha$ and phases $\theta_\alpha$ of the incident waves injected into the system from 
the $\alpha-$th TL. These wavefront shaping parameters are encoded in {\bf b}; resulting in ${\displaystyle 
\frac{\partial b_n}{\partial A_\alpha} = 2 i e^{i \theta_\alpha} W_{n,\alpha}}$, and ${\displaystyle \frac{\partial b_n}{\partial 
\theta_\alpha} = -2 A_\alpha e^{i \theta_\alpha} W_{n,\alpha}}$.

Eventually, the objective function gradient becomes:
\begin{equation}
    \frac{d g}{d \mathbf{p}} \equiv \left[ \frac{d g}{d L_{nm}^{\rm opt}},\ \frac{d g}{d A_\alpha},\ \frac{d g}{d \theta_\alpha} \right]=
\begin{bmatrix}
        \displaystyle 2 \mathcal{R} \left\{ \Psi^\dagger \frac{\partial {\cal M}}{\partial L_{nm}^{\rm opt}} \Phi \right\} \\[2ex]
        \displaystyle \frac{\partial g}{\partial A_\alpha} + 2 \mathcal{R} \left\{ \Psi^\dagger \frac{\partial b}{\partial A_\alpha} \right\} \\[2ex]
        \displaystyle \frac{\partial g}{\partial \theta_\alpha} + 2 \mathcal{R} \left\{ \Psi^\dagger \frac{\partial b}{\partial \theta_\alpha} \right\}
    \end{bmatrix}^\top.
    \label{grad_graph}
\end{equation}
It is important to emphasize that Eq. (\ref{grad_graph}) does not require measuring the entire $\Phi$ or $\Psi$ but only those voltages 
that correspond to the non-zero entries of ${\partial{\cal M}\over \partial L_{nm}^{\rm opt}}$ and ${\partial {\bf b}\over \partial 
A_\alpha}$, ${\partial {\bf b}\over \partial \theta_\alpha}$ associated with the controllable parameters. In other words, the sparsity 
of $\partial M/\partial{\bf p}$ and $\partial{\bf b}/\partial{\bf p}$ further reduces measurement complexities.

\section{Examples of Modalities Associated with Specific Optimization Objective Functions}

Below we provide some examples of optimization objective functions associated with various modalities.

{\it Targeted Mode Transfer --} In many practical scenarios, particularly in in-door wireless communications, one requires energy/information 
transfer from specific input channels to designated output channels -- distinct from the injected ones. Such targeted mode transfer (TMT) can be achieved by an appropriate cavity-shaping and/or wavefront-shaping process whose success is quantified by the objective 
function
\begin{equation}
    g_{\text{TMT}} = \frac{\sum_{\{T_\alpha\}} |\phi_\alpha|^2}{\sum_{\{I_\beta\}} |A_{\beta}|^2},
\end{equation}
where $\{T_\alpha\} (\neq \{I_\beta\})$ denotes the set of targeted (injected) channels. In case of lossless structures $g_{\text TMT}=1 (0)$ 
indicates perfect (poor) TMT performance.

{\it Coherent Perfect Absorption --} Coherent perfect absorption (CPA) \cite{aluCPA,CaoCPA2011,chen2020perfect} requires that the incident 
radiation has a particular frequency and spatial field distribution (coherent illumination) such that the (weakly) absorbing cavity acts as 
a perfect constructive interference trap that eventually absorbs completely the incident radiation. The adjoint optimization methodology can 
be utilized for the management of the multi-path constructive interference via cavity shaping and/or wavefront shaping. In this case, the 
optimization objective function is
\begin{equation}
    g_{\text{CPA}} = 1 - \frac{\sum_{\{I_\alpha\}} |\phi_\alpha - A_\alpha e^{i \theta_\alpha}|^2}{\sum_{\{I_\alpha\}} |A_\alpha|^2} - 
    \frac{\sum_{\{T_\beta\}} |\phi_\beta|^2}{\sum_{\{I_\alpha\}} |A_\alpha|^2}
\end{equation}
where the second term describes the reflected waves from the injected channels $\{I_\alpha\}$ and the third term describes the transmitted 
waves from the remaining $\{T_\alpha\}\neq \{I_\beta\}$ channels. Perfect absorption corresponds to $g_{\text{CPA}}=1$.

{\it Invisibility --} Evading the detectability of a scattering object requires the elimination of any imprints in the phase and amplitude 
of the scattered interrogating waves due to their interaction with a target. This is achieved by appropriate manipulation of the many-path 
interference phenomena occurring inside the scattering domain via cavity shaping and/or tailoring control signals that counter phase and 
amplitude scattering imprints (including absorption) caused by interactions with the target. The objective function that ensures such optimal 
cancellations take the form
\begin{equation}
    g_{\text{invis}} = \frac{|\phi_{\alpha_0} - A_{\beta_0} e^{i \theta_{\beta_0}}|^2}{A_{\beta_0}^2}+  
    \frac{|\phi_{\beta_0} - A_{\beta_0} e^{i\theta_{\beta_0}}|^2}{\sum_{\{I_\beta\}} |A_\beta|^2}+ 
    \frac{\sum_{\{T_{\alpha\neq \alpha_0,\alpha_c}\}} |\phi_\alpha|^2}{\sum_{\{I_\beta\}} |A_\beta|^2},
    \label{transparent}
\end{equation}
where $g_{\rm invis}=0$ indicates optimal invisibility/camouflage performance. Above, the first term on the right-hand-side compares the 
scattered signal (phase and amplitude) from a probed $\alpha_0-$channel to an interrogating signal injected into the system from a $\beta_0$
-channel; the second term measures the reflectance from the $\beta_0$-channel. Finally, the last term evaluates the transmittance to all 
channels that are different from the probe channel $\alpha_0$ and the control channel $\alpha_c$. The objective function Eq. (\ref{transparent}) 
does not enforce any constraints to the reflected wave from the control channel $\alpha_c$.

\section{In-situ Implementation of iPAC}

\begin{figure*}
\centering
\includegraphics[width=\linewidth]{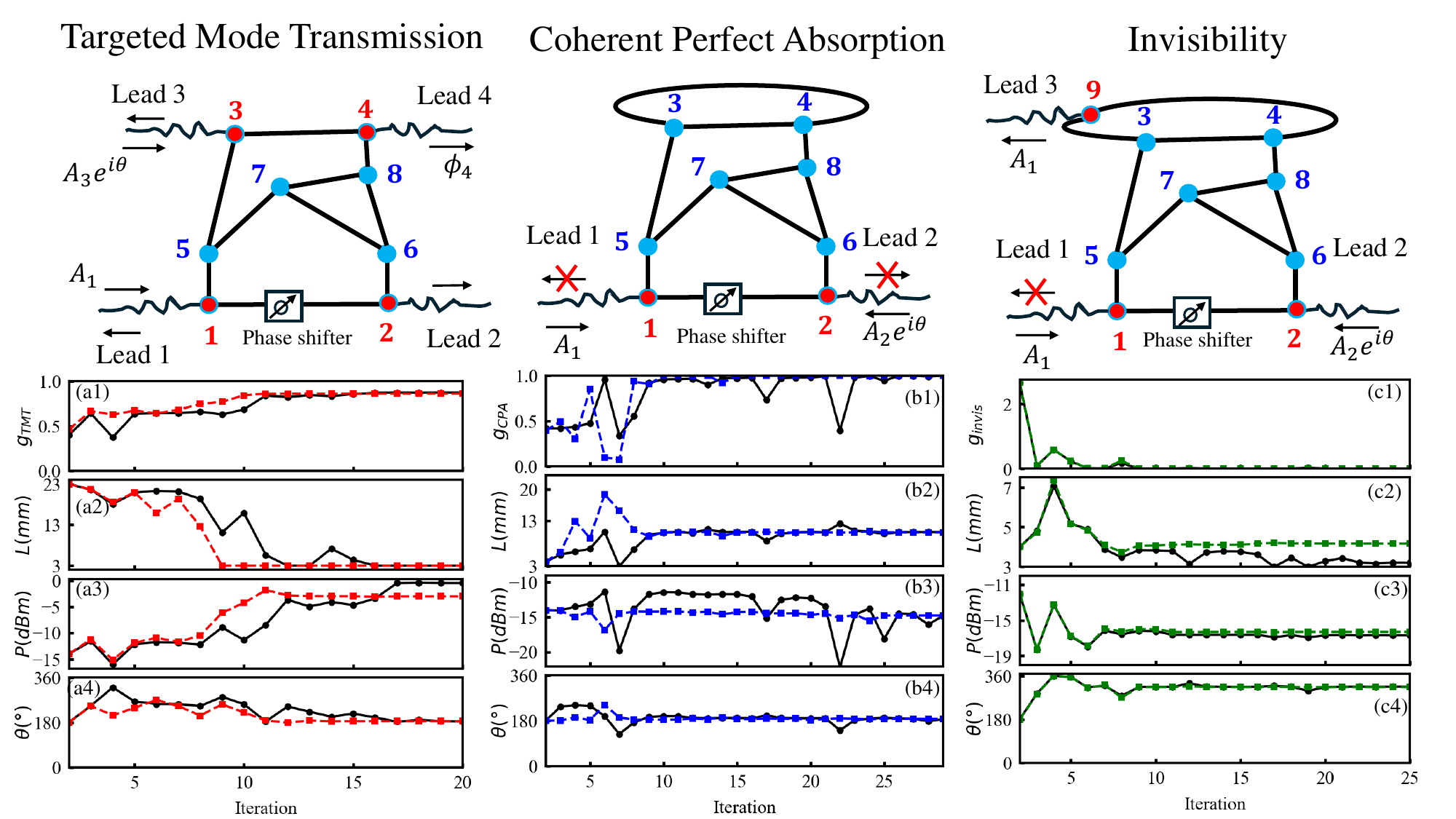}
\caption{\label{Fig3}%
{\bf Experimental demonstration of IPAC optimizer:} {\it In-Situ} demonstration of the iPAC optimizer using a complex network of coaxial 
cables (see upper row for network schematics). The red vertices indicate the positions where a TL is attached. The black solid (colored 
dashed) lines with filled black circles (colored squares) are the experimental (digital twin) results. The control parameters used for 
the optimization involve the (electrical) length of the coaxial cable $L_{12}^{\rm opt}$ (using the attached phase shifter) and the 
relative phase and amplitude of the injected signals. Three different modalities are demonstrated: \textbf{First column} 
(\textbf{a1--a3}): Targeted Mode Transmission (TMT) where the injected wavefront (with frequency $f=1.86$GHz) from TLs $\alpha=1,3$ and the 
cable length $L_{12}^{\rm opt}$ are optimized to deliver the input signal to the targeted TL $\alpha=4$ with efficiency $\sim 87\%$. (a1) 
Convergence of the TMT objective function \(g_{\mathrm{TMT}}\) vs. iteration number. Evolution of (a2) the electrical length $L_{12}^{\rm opt}$; 
(a3) the injected relative power and (a4) relative phase (with respect to a signal injected from TL $\alpha=1$) of the signal injected from 
TL $\alpha=3$. \textbf{Second column} (\textbf{b1--b3}): Coherent Perfect Absorption (CPA) for a wave injected from TLs $\alpha=1, 2$ at 
frequency $f=3.26$GHz. (b1) Convergence of the CPA objective function \(g_{\mathrm{CPA}}\) toward nearly perfect absorption (\(g_{\mathrm{CPA}} 
\approx 0.9998\)). Evolution of (b2) the electrical length $L_{12}^{\rm opt}$; (b3) injected relative power and (b4) relative phase (with 
respect to a signal injected from TL $\alpha=1$) of the signal injected from TL $\alpha=2$ vs. iteration number. \textbf{Third column} 
(\textbf{c1--c3}): Signal invisibility (cavity-camouflage). The interrogating signal at frequency $f=0.74$GHz is injected into the network 
from TL $\beta_0=1$ and is received from TL $\alpha_0=2$ with the same amplitude and phase ($0.01$dB power variation and $0.1^o$ phase variation 
with respect to the injected wave). A control signal (phase and amplitude) injected from channel $\alpha_c=3$ is balancing the losses and 
together with the length $L_{12}^{\rm opt}$ ensures the invisibility of the cavity as far as the processing signal at TL $\alpha_0=3$ is 
concerned. The reflected signal from TL $\beta_0=1$ is essentially zero. (c1) Convergence of \(g_{\mathrm{invis}}\) to \(\sim  10^{-4}\), 
signifying that the transmitted field matches the desired (injected) wave. Evolution of (c2) the cable-length $L_{12}^{\rm opt}$; (c3) injected 
relative power; and (c4) relative phase (with respect to a signal injected from TL $\alpha=1$) of the control signal injected from TL $\alpha_c=3$ 
vs. iteration number.
}
\end{figure*}

We proceed with the {\it in-situ} implementation of our optimization scheme for the three modalities discussed above. The schematics of the 
microwave networks for each of the three cases are shown in the upper row of Fig. \ref{Fig3}. The TLs (black wiggling lines) are attached 
to vertices that are indicated with red-filled circles. The amplitude and phase of the injected signals from a two-source VNA, have been 
used as optimization parameters (wavefront-shaping). In all cases, the signal from the $\alpha=1$ TL serves as a reference for the amplitude 
$A$ and the phase $\theta$ of the signal injected from the second TL. Finally, the bond $L_{12}^{\rm opt}$ incorporates a phase-shifter which 
was digitally controlled for ``cavity-shaping'' purposes.

The in-situ optimization protocol proceeds as follows: {\bf (a)} {\it Forward Measurement:} First, we inject signals from TLs attached to 
two of the vertices ($n=1,2$ for the CPA and $n=1,3$ for the TMT and invisibility modalities, see upper row of Fig. \ref{Fig3}) into the 
network. Forward voltages $\Phi$ are measured at the vertices $n=1,2$ that are associated with the length optimization parameter $L_{12}^{\rm 
opt}$. {\bf (b)} {\it Adjoint Measurement:} The source for the adjoint measurement is then constructed from the previously measured forward 
voltages and according to the specific objective function. For the TMT case, the adjoint input was delivered from the $\alpha=4$ TL that 
was targeted for maximizing the outgoing signal. For the CPA protocol, the adjoint input was delivered to the two $\alpha=1,2$ TLs. Finally, 
for the invisibility protocol, the adjoint signal was delivered to $\alpha=1,2$ TLs where the optimization constraints have been imposed. 
In this case, a control field was also injected from the remaining $\alpha=3$ TL. The adjoint 
voltages $\Psi$ that were needed to measure for the evaluation of the gradient were associated with the vertices connected to the TLs that 
have been used to inject the input signal, i.e. $\alpha=1,2$ for the CPA, $\alpha=1,3$ for the TMT and invisibility protocols; {\bf (c)} 
{\it Gradient Calculation:} With the {\it local} forward $\Phi$ and adjoint $\Psi$ measurements obtained, we calculate the gradient of the 
objective function with respect to the controllable parameters (see Eq. (\ref{grad_graph})) in real-time. We re-emphasize that the choice 
of controllable parameters dictates the positions/vertices $n$ where the required forward $\Phi_n$ and adjoint measurements $\Psi_n$ of the 
voltages are performed for the gradient calculation $\frac{dg}{dp}$; {\bf (d)} {\it Parameter Update:} Once the gradient was computed, we 
employed a gradient descent algorithm to update the optimization parameters. Specifically, we used the package NLOPT with the Limited-memory 
Method of Moving Asymptotes (LD\_MMA) option \cite{LDMMA}. This algorithm identifies a new set of parameters, which were then implemented 
by adjusting the phase shifter length and modifying the relative amplitude and phase of the input signals; {\bf (e)} {\it Iteration and 
Convergence:} The steps (a-d) are repeated until the objective function converges on an optimal value (within some tolerance). A single 
operating frequency was selected and held constant throughout the optimization process. 

In Fig. \ref{Fig3} we report the results of the in-situ optimization (solid black lines with filled circles) for the three modalities discussed 
above. In all cases, we have achieved a rapid convergence towards an optimal value of the corresponding objective functions occurring after 
$\sim 20$ iterations of the protocol, see Figs. \ref{Fig3}(a2,b2,c2). The convergence of the three control parameters ($L_{12}^{\rm opt}, A, 
\theta$) towards their optimal value for each of the three modalities is reported in the third, fourth, and fifth rows of the same figure 
respectively.

\section{In-silico Implementation of iPAC for large control parameter system}

\begin{figure*}
\includegraphics[width=0.9\linewidth]{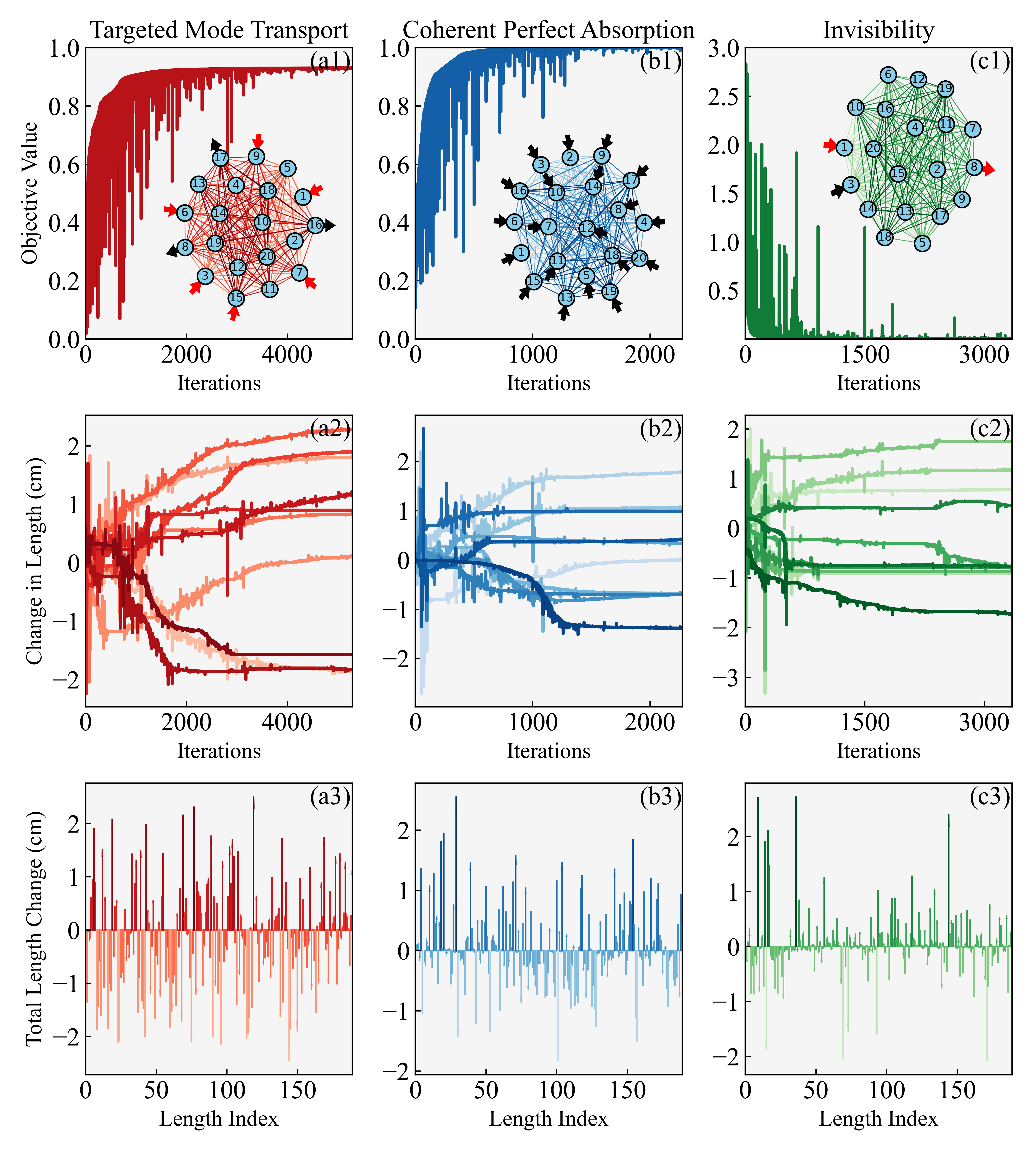}
\caption{ \label{Fig2}
{\bf In-silico demonstration of the iPAC optimizer:} In-silico demonstration of the iPAC scheme using a digital twin of a fully connected 
network of \(V=20\) vertices consisting of $190$ lossy coaxial cables. Each vertex is attached to a TL. The control parameters used for 
the in-silico optimization involve only the bonds of the network (cavity shaping). Three different modalities, all at $f=3.2$GHz, are 
demonstrated: \textbf{First column} (\textbf{a1--a3}): Targeted Mode Transmission (TMT) for a scenario where six TLs $\alpha=1, 3, 6, 7, 
9,$ and $15$ are used to inject a random wavefront, and the network is optimized to deliver the input signal \( \approx 90\% \) 
to the targeted channels $n=8, 16, 17$. (a1) Convergence of the TMT objective function \(g_{\mathrm{TMT}}\) vs. iteration number. (a2) 
Representative evolution of selected bond-length variations during the optimization. (a3) Final set of bond-length variations across all 
network bonds. \textbf{Second column} (\textbf{b1--b3}): Coherent Perfect Absorption (CPA) scenario for a wave injected from all \(N=20\) 
TLs. (b1) Convergence of the CPA objective function \(g_{\mathrm{CPA}}\) toward nearly perfect absorption (\(g_{\mathrm{CPA}} \approx 
0.9998\)). (b2) Representative bond-length variations vs. iteration. (b3) Final network configuration achieving the CPA state. \textbf{
Third column} (\textbf{c1--c3}): Signal invisibility (cavity-camouflage). The interrogating signal is injected into the network from 
channel $\beta_0=1$ and is received from TL $\alpha_0=8$ with the same amplitude and phase. A control signal injected from a control channel 
$\alpha_c=3$ is balancing the losses. (c1) Convergence of 
\(g_{\mathrm{trans}}\) to \(\sim 6 \times 10^{-4}\), signifying that the transmitted field matches the desired (injected) wave. (c2) Evolution 
of selected bond-length variations during the optimization. (c3) Final distribution of bond-length variations across all bonds after 3351 iterations.
}
\end{figure*}

In Fig. \ref{Fig2}, we also present the {\it in-silico} results (shown as dashed lines with filled squares) obtained from a digital twin 
implementation of the AO protocol. The close agreement between the digital twin and the \textit{in-situ} results confirms that our experimental set-up is adequately captured by our network model. This validation supports the applicability of the digital twin approach to more complex networks, with a larger number of control parameters.

We considered fully connected networks consisting of $N=20$ vertices with a total of $190$ bonds. The bond-lengths are initially uniformly 
distributed in the interval $[L_0-\delta L, L_0+\delta L]$ where $L_0=25$cm and $\delta L=5$cm. At each vertex, we have attached TLs (i.e. 
$N=20$) that have been used for injecting (receiving) the interrogating (scattering) signal. The frequency of the injected waves was chosen 
to be $f=3.2$GHz for all cases. The optimization process has been achieved via bond-length variations (cavity-shaping approach).

The first column of Fig. \ref{Fig2} reports the results of the TMT modality. A random wavefront has been injected into the network from 
six TLs coupled to vertices $n=1, 3, 6, 7, 9,$ and $15$. The adjoint optimization scheme aimed to identify the appropriate bond-length 
variations that resulted in delivering the injected signal to a specified set of channels attached to vertices $n=8, 16, 17$, see inset 
of Fig. \ref{Fig2}(a1). In the main part of subfigure Fig. \ref{Fig2}(a1) we show the convergence of the objective function $g_{\rm TMT}$ 
to a total transmittance of approx. $90\%$. We have checked that for the converged optimal bond-length configuration, the remaining $10\%$ 
energy loss was associated with the absorption due to Ohmic losses at the wires. Typical bond-length variations $\delta L_{nm}^{\rm opt}$ 
versus the iteration number are shown in Fig. \ref{Fig2}(a2), while in Fig. \ref{Fig2}(a3) we report the bond-length variations for 
all bonds of the network at the end of the optimization process. 

The second column of Fig. \ref{Fig2} reports the digital twin calculations for the CPA-scenario. We have injected a randomly chosen coherent 
wavefront from all $N=20$ TLs into the lossy network, see inset of Fig. \ref{Fig2}(b1). Using the adjoint optimization protocol we have 
determined the optimal bond-lengths for which the network acts as a perfect constructive interference trap leading to complete absorption 
of the incident wave. In Fig. \ref{Fig2}(b1), we show the convergence of the objective function towards a value $g_{\rm CPA}= 0.9998$ occurring 
after 2276 iterations. The evolution of some typical bond variations versus the number of iterations is shown in Fig. \ref{Fig2}(b2), 
while Fig. \ref{Fig2}(b3) reports the final bond-variation for all bonds.

Finally, the last column of Fig. \ref{Fig2} shows the digital twin simulations for cavity camouflage (invisibility), see the inset 
of Fig. \ref{Fig2}(c1). We have injected a signal from channel $\beta_0=1$ with an amplitude $A_{\beta_0}=0.56$ and phase $\theta_{
\beta_0}=69^o$. To balance the network losses we have also injected a control signal into the system from channel $\alpha_c=3$ with amplitude 
$A_{\alpha_c}=0.85$ and phase $\theta_{\alpha_c}=333^o$. The adjoint optimization protocol aimed to identify an appropriate bond-length 
configuration for which the scattered signal collected at a specified $\alpha_0=8$ TL is identical to the interrogating wave injected from 
TL $\beta_0=1$. In Fig. \ref{Fig2}(c1) we show the convergence of the objective function $g_{\rm trans}$ towards the value $g_{\rm invis}
\approx 6\times 10^{-4}$ after $3351$ iterations. Here, in addition to the constraints imposed by the objective function $g_{\rm invis}$ 
in Eq. (\ref{transparent}) we have also requested zero transmission and reflection from the control channel $\alpha_c=3$. This additional 
constraint introduce the following 
modification $g_{\rm invis}\rightarrow g_{\rm invis} +\frac{|\phi_c-A_{\alpha_c}e^i\theta_{\alpha_c}|^2}{\sum_{\{I_\beta\}} |A_\beta|^2}$. Even 
with a modest number of iterations $\sim 400$ of the adjoint optimization scheme the objective function can be as small as $g_{\rm trans}\approx 
0.01$. Figures \ref{Fig2}(c2,c3) show a representative evolution of bond-length variations versus iteration number and the final bond-lengths 
at the end of the optimization process ($3351$ iterations).

\section{Discussion}

In conclusion, we have presented a proof-of-concept experimental demonstration of \textit{in-situ} Physical Adjoint Computing for real-time control 
of electromagnetic wave modalities in complex multi-resonant and multi-scattering systems. Our approach can be mapped onto in-door wireless communication 
protocols~\cite{karamanos2024topology}, in which a reconfigurable intelligent surface can be swiftly programmed to deliver stronger signals to moving targets amidst an evolving 
environment in real time. In such protocols, electric field measurements need to be made only at the positions of the metasurface elements and the 
target. Importantly, no knowledge is required of the full electromagnetic environment, including any big or small obstacle which may stand in the 
way or even moving. Therefore, our approach is fundamentally different from existing wavefront-shaping methodologies, which require a full knowledge 
of the scattering matrix and its eigen-decomposition to identify the optimal wavefront patterns for achieving specific operations. Crucially, in 
such methods, the entire scattering matrix needs to be repeatedly re-measured and re-analyzed every time the metasurface is reconfigured and/or the 
surrounding changes, leading to formidable challenges in larger and more complex environments. In contrast, in-situ adjoint optimization bypasses 
the need for a scattering matrix by directly exploiting the gradient sensitivities judiciously plucked from a set of strategically positioned 
measurements.   

Our wave-network platform also significantly differs from physical implementations of feed-forward neural networks, which typically do not utilize 
complex (multi-scattering) wave interactions in a multi-resonant electromagnetic environment. By leveraging these interactions, our platform amplifies 
small variations in the optimization parameters via multiple interference pathways, deriving richer physical abilities from a relatively smaller number 
of controllable parameters (in contrast to billions of weights and biases required in a feed-forward neural net). While we do not pursue any deep 
learning functionality in this work (and thus require no training data), we note that our setup offers a natural ``physics-aware'' deep learning 
platform for both in-situ training and inference, rather than a cumbersome imitation of an abstract neural network architecture. Most importantly,
our experiments pave the way for the development of more powerful in-situ optimization protocols which will involve nonlinear and non-reciprocal 
wave mechanics, broadband pulses, and real-time control learning.

{\it Acknowledgements --} We acknowledge partial funding from MURI ONR-N000142412548, DOE DE-SC0024223, NSF-RINGS ECCS (2148318), Simons
Foundation (SFI-MPS-EWP-00008530-08), and US Army Research Office (ARO) W911NF2410390.

\newpage
\clearpage
\appendix

\section{Methods}
\subsection{Network modeling}
The transport properties of the microwave network are modeled using a metric graph consisting of one-dimensional wires (bonds) supporting a single 
propagating mode. The waves propagating between the inner and the outer conductor along the 
coax cable, is given in terms of the difference $\psi_{B}(x_B)$ between the potentials at the conductors' surfaces, see Eq. (\ref{telegraph}). The 
bonds are connected together at vertices ($v$-port dividers) where Neumann boundary conditions are imposed. In the experimental network we have used 
$3-port$ vertices (T-junctions). 

The solutions of Eq. (\ref{telegraph}) at a bond $B=(nm)$ can be expressed as
\begin{equation}
    \psi_{B}\left( x_{B} \right) = \phi_n\, \frac{\sin\left[ k\left( L_{B} - x_{B} \right) \right]}{\sin\left( k L_{B} \right)} + \phi_m\, 
    \frac{\sin\left( k x_{B} \right)}{\sin\left( k L_{B} \right)}
    \label{wfg}
\end{equation}
which satisfy the wave continuity conditions $\psi_{B}\left( x_{B} = 0 \right) = \phi_n; \psi_{B}\left( x_{B} = L_B \right) = \phi_m$, for each 
pair of connected vertices $n<m$. Furthermore, the current is conserved at each vertex, i.e.,
\begin{equation}
    \sum_{m} \left. \frac{d\psi_{B}\left( x_{B} \right)}{dx_{B}} \right|_{x_{B} = 0} + \sum_{\alpha' = 1}^L \delta_{\alpha\alpha'} \left. 
    \frac{d\psi_{\alpha'}(x)}{dx} \right|_{x = 0} = 0,
\end{equation}
where $\delta_{\alpha\alpha'}$ is the Kronecker delta, and the second term accounts for the derivatives at the ports connected to TLs. Combining the 
above vertex boundary conditions, together with Eq. (\ref{wfg}) we arrive at Eq. (\ref{grapheqmot}) which provides the system matrix that describes 
transport in the forward direction.


\subsection{In-silico Implementation of the iPAC}

We conducted in-silico simulations of a complex scattering network consisting of $V=20$ vertices that are fully connected with $190$ bonds. Each 
vertex was attached to a TL, i.e., $N=20$. In the digital twin simulations, we considered a cavity-shaping optimization scheme that allows adjustments 
of all $190$ bond-length, i.e., a large number of DoF relevant to operational realities. In fact, under such conditions is expected that the 
implementation of the adjoint-based gradient descent protocol is more beneficial as compared to other optimization schemes

The bond lengths were initialized with randomized values uniformly distributed around $L_{\text{0}} = 0.25$~meters, with variations constrained to 
$\delta L\pm 5$cm ($=0.5\lambda$ where $\lambda$ is the operational wavelength), such that $L_{nm} \in \left[ L_{0} - \delta L,\, L_{0} + \delta 
L\right]$. For all modalities, the wavefront parameters were fixed, with the amplitude parameters constrained $A_i \in [0.001,\, 3.0]$, and 
$\theta_i \in [-\pi,\, \pi]$. An amplitude lower bound was set to prevent trivial solutions with zero input power. 

\subsection{Controlling and setting up the VNA for coherent wavefront shaping inputs}

To precisely control the phase and amplitude of signals injected into the input ports of the scattering system, a Keysight PNA P5023B four-port Vector 
Network Analyzer (VNA) equipped with the S93089B Differential and I/Q Device Measurements option was utilized. The S93089B option enables accurate phase 
control of multiple internal sources, facilitating coherent excitation without the need for external hybrid couplers or baluns. Two internal sources 
were configured to deliver signals to the desired input ports of our microwave graph network. Both sources were set to the same frequency to maintain 
coherence, while the relative phase between the two sources was precisely adjusted from $0^o$ to $360^o$ using the S93089B's phase control settings. This 
allows the phase difference to be fixed at specific values in degrees of one input port relative to a reference port. The output power of each source  
was individually adjusted in the interval $[-40dBm, 0dBm]$ to achieve the desired amplitude difference at each input port. A calibration routine was 
executed to compensate for any 
inherent phase and amplitude imbalances introduced by the VNA's internal signal paths and external cabling. For ensuring the experimental stability of 
the objective function, we performed ten measurements per iteration, resulting in essentially identical outputs characterized by their mean value.  

For forward scattering measurements, both sources were activated and phase-aligned according to a random set of initial values. The S93089B option's 
source-phase control ensured that the relative phase between the inputs was maintained with high precision throughout the measurement. The VNA's receivers 
were configured to measure the DUT's response at the fundamental frequency, capturing the effects of coherent excitation on the forward scattering 
parameters. In adjoint measurement scenarios, these sources provided the excitation signal, and the VNA measured the reflected and transmitted signals 
accordingly (in case only one excitation signal was needed the other source was deactivated).

\subsection{Controlling and setting up the mechanical phase shifter}

To achieve precise length manipulation in our experimental setup, we integrated a mechanical phase shifter into the system (bond $L_{12}^{\rm opt}$), 
a coaxial RF phase shifter, typically operated via a manual knob. To enable automated and repeatable control, we motorized the phase shifter and 
developed a characterization method to correlate motor movements with the resulting length perturbations. 

The mechanical phase shifter used in our experiment was designed to operate over a frequency range of DC to 18 GHz, with an insertion loss of less than 
$1.0$ dB up to $18$ GHz and capable of handling up to $100$ Watts of average RF power. The device originally featured a manual adjustment knob for phase 
tuning, however, a stepper motor was mechanically coupled to the phase shifter's adjustment knob. The motor was securely mounted to maintain alignment 
and prevent mechanical backlash, ensuring consistent control over the phase adjustment mechanism. The motor was interfaced with a Trinamic motion controller, allowing for precise digital control of the motor's position, and providing an adjustable phase shifter's length from $3m$m to $23m$m. The controller 
was connected to a computer via a USB interface, enabling automated control through custom Python scripts.

To establish a reliable relationship between the motor's rotational steps and the physical displacement within the phase shifter, we conducted a calibration 
process using a high-precision digital micrometer (Asimeto IP65 Digital Outside Micrometer). The micrometer was positioned to measure the linear displacement
resulting from the motor's rotation. The micrometer's spindle was placed in contact with a reference point on the phase shifter that moved in response 
to the internal adjustment mechanism. The motor was programmed to move in increments of microsteps, and the corresponding displacement was recorded using 
the micrometer. Movements were performed in both clockwise and counterclockwise directions to account for any mechanical hysteresis. The collected data 
indicated that 464 microsteps of the motor corresponded to a linear displacement of 1 mm within the phase shifter. A linear relationship was established 
between the number of micro-steps ($N_{steps}$) and the displacement $d = \frac{N_{steps}}{464}m$m. Multiple trials were conducted to confirm the repeatability 
of the calibration. The standard deviation of the displacement measurements was within the micrometer's specified accuracy, ensuring confidence in the 
calibration.

\subsection{Modeling of the phase shifter}

To understand the phase shifter's impact on the transmitted signals, we modeled it as a variable-length transmission line supporting a Transverse 
Electromagnetic (TEM) mode. The phase shift introduced by the device is a function of the electrical length, which depends on both the physical 
length and the dielectric properties of the medium. Using the calibrated Vector Network Analyzer (VNA) setup described previously, we measured the 
scattering matrix ($S$-parameters) of the phase shifter over the frequency range of interest. Measurements were taken at various positions of the phase 
shifter corresponding to different micrometer readings. The phase shifter was modeled as a two-port network with its behavior represented by transmission 
line equations. The phase shift ($\phi$) introduced by the line is given by $\phi = \beta \cdot d$ where $\beta$ is the phase constant, and $d$ is 
the physical length of the transmission line. The phase constant is related to the frequency $f$ and the effective permittivity $\epsilon_{eff}$ of 
the medium by $\beta = \frac{2\pi f}{c}\sqrt{\epsilon_{eff}}$ where $c$ is the speed of light in a vacuum. The effective permittivity was assumed 
to be complex to account for dielectric losses within the phase shifter. We modeled $\epsilon_{eff}$ as a function of the micrometer-measured length 
and frequency. To extract the relationship between the effective permittivity, physical displacement, and frequency, we employed surrogate optimization 
using MATLAB finding that $n = \epsilon_{eff} \approx 1.004 + 0.0022i$ and the functional dependence of the length of the phase shifter 
$L_S$ on the measured length of the micrometer $L_{ps} = 286mm+2\left(d-7mm\right)$ accounting for the fact that it is a trombone line phase shifter, 
so the factor of 2 accounts for the doubling of the line when making length adjustments.

\bibliography{adjoint_invivo}
\end{document}